\documentclass[aps,twocolumn,prb,preprintnumbers,amsmath,amssymb]{revtex4-2}
\allowdisplaybreaks[4]
\newcommand{\lsim} 
 {\ \raise.35ex\hbox{$<$}\kern-0.75em\lower.5ex\hbox{$\sim$}\ }
\newcommand{\gsim}
 {\ \raise.35ex\hbox{$>$}\kern-0.75em\lower.5ex\hbox{$\sim$}\ }

\usepackage{graphicx}
\usepackage{dcolumn}
\usepackage{bm}
\usepackage{amsmath}
\usepackage{times}
\usepackage[usenames]{color}
\usepackage{natbib}
\usepackage{ulem}

\begin{document}
\title{Altermagnetic Perovskites}
\author{Makoto Naka$^{1*}$, Yukitoshi Motome$^2$ and Hitoshi Seo$^{3,4}$}
\affiliation{$^1$School of Science and Engineering, Tokyo Denki University, Ishizaka, Saitama 350-0394, Japan}
\affiliation{$^2$Department of Applied Physics, The University of Tokyo, Bunkyo, Tokyo 113-8656, Japan}
\affiliation{$^3$Condensed Matter Theory Laboratory, RIKEN, Wako, Saitama 351-0198, Japan}
\affiliation{$^4$Center for Emergent Matter Science (CEMS), RIKEN, Wako, Saitama 351-0198, Japan}
\email{m-naka@mail.dendai.ac.jp}
\date{\today}
\begin{abstract}
Altermagnet is a class of antiferromagnets, which shows a staggered spin ordering with wave vector ${\bm q}=0$, while its net magnetization is cancelled out in the limit of zero relativistic spin-orbit coupling. 
The simplest case is when the up and down spins are ordered on two crystallographically equivalent sublattice sites within the unit cell that are not connected by translation, and consequently, the system breaks the macroscopic time-reversal symmetry.
Accordingly, it exhibits non-relativistic spin splitting in the energy band and characteristic cross-correlation phenomena between spin, charge, and lattice (orbital) degrees of freedom. 
This is in contrast to conventional N\'{e}el-type antiferromagnets with ${\bm q} \neq 0$ conserving the macroscopic time-reversal symmetry, where the time-reversal operation flipping of spins combined with translation can make the system identical to the original state. 
Altermagneticsm is universally latent in various magnetic materials that have been considered as simple collinear-type antiferromagnets.
In this article, we focus on perovskites with chemical formula {\it ABX}$_3$, which are typical playgrounds for strongly correlated electron systems, and overview their altermagnetic aspects that have been overlooked in the past researches, based on microscopic model studies revealing the mechanisms of their properties.
We display that a combination of a variety of antiferromagnetic ordering and the commonly-seen lattice distortions in perovskites gives rise to a non-relativistic spin splitting whose mechanism does not rely on the spin-orbit coupling and its consequent spin current generation, and the anomalous Hall effect in the presence of the spin-orbit coupling. 
\end{abstract} 


\maketitle
\narrowtext



%
%

%


\section{Introduction\label{sec:intro}}

Perovskites, a large family of compounds with chemical formula {\it ABX}$_3$ 
as their mother phase, 
show versatile physical properties and serve as one of the most well-studied textbook materials in condensed matter physics~\cite{tilley}.
In particular, transiton-metal-based (TM-based) perovskites exhibit a wide range of functional properties~\cite{imada,cheong,maekawa}, 
e.g., ferroelectricity, metal-to-insulator (MI) transition, magnetoelectric effect, spin crossover phenomenon, superconductivity, and photovoltaic effect.
In recent years, a new chapter has been added to 
them~\cite{Naka_PRB2021,Naka_PRB2022}: 
``altermagnetism''. 
\begin{figure}[t]
\begin{center}
\includegraphics[width=0.8\columnwidth, clip]{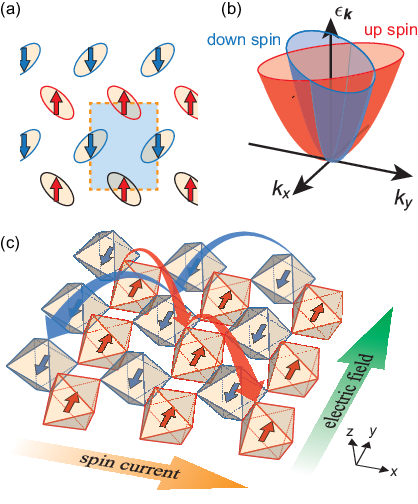}
\end{center}
\caption{
(a) A minimal altermagnet, where local spin moments locate on 
different anisotropic electronic orbitals, represented by ellipses, which can be owing to oriented molecules or ligand structures.
The shaded rectangle denote the unit cell common for the crystal structure and the magnetic order (${\bm q}=0$).
(b) A $d$-wave-type spin-split band structures for up and down spin electrons in an altermagnet.
(c) Schematic illustration of altermagnetic perovskites. 
The red and blue curved arrows represent the sublattice-dependent spatially anisotropic hoppings of up and down spin electrons, respectively, that render perovskite as altermagnet.
}
\label{fig1}
\end{figure}

Altermagnets~\cite{Smejkal_PRX2022} refer to magnetic materials which appear to be conventional collinear N\'{e}el-type antiferromagnets but actually break the time-reversal (TR) symmetry due to the lattice structure behind the antiferromagnetic (AFM) order. 
This happens in a wide variety of materials, whose schematic view is shown in Fig.~\ref{fig1}(a)~\cite{Naka_NatComm2019}.
The cause of TR symmetry breaking is that the two sites where up and down spins exist are not connected by inversion or any translation operation, while they are connected by mirror, glide, or screw operation.
Therefore, the net magnetization is cancelled out, in the limit of zero relativistic spin-orbit coupling (SOC)~\cite{net_magnetization}.
In this situation, unlike the local TR symmetry breaking in conventional antiferromagnets, here the macroscopic TR symmetry is broken~\cite{PT_symmetry}.
Consequently, altermagnets show cross-correlation phenomena unexpected in conventional antiferromagnets but rather reminiscent of ferromagnets. 
Therefore, in contrary to N\'{e}el's statement in his Nobel lecture on antiferromagnets: {\it They are extremely interesting from theoretical viewpoint, but do not seem to have any application}~\cite{Neel_nobel}, now they are attracting much attention as antiferromagnets with possible applications.

A notable property of altermagnets is the non-relativistic spin splitting appearing in their band structure, 
 resulting in a novel mechanism of spin current generation, 
independently proposed by Ahn {\it et al.} in a rutile-type oxide RuO$_2$~\cite{Ahn_PRB2019} 
 and Naka {\it et al.} in an organic antiferromagnet $\kappa$-(ET)$_2X$~\cite{Naka_NatComm2019} in 2019.
Figure~\ref{fig1}(b) shows the schematic spin-split band structure in altermagnets, which occurs even in the absence of the SOC, in stark contrast to the relativistic spin momentum locking, e.g., the Rashba effect. 
As discussed in Ref.~\cite{Naka_NatComm2019} based on the Hubbard model,  
the splitting originates from a cooperative effect of an AFM ordering and spatially anisotropic electron hoppings that are sublattice dependent, which can also be interpreted as an AFM analog of the exchange splitting in ferromagnets. 
The resultant $d$-wave like anisotropic spin splitting in the Brillouin zone enables us to generate a spin current under an external field.

Another feature of altermagnets is the anomalous Hall effect (AHE) 
 proposed by Smejkal {\it et al.}, also investigating RuO$_2$ in 2020~\cite{Smejkal_SciAdv2020,note_arXiv}, 
 which is a relativistic effect based on the SOC. 
Compared to a numerous number of studies on the AHE in non-coplaner and non-collinear antiferromagnets in the last decade~\cite{Nagaosa_RMP2010, Smejkal_NatRevMat2022}, 
 as we will discuss below, the AHE in altermagnets is characterized by the TR and mirror symmetry breaking not only due to the AFM spin pattern but also the background lattice structure~\cite{Naka_PRB2020}. 
Since then, enormous amount of papers have been published in a short period~\cite{Hayami_JPSJ2019, Hayami_PRBL2020, Hayami_PRB2020, Yuan_PRB2020, Yuan_PRB2021, Samanta_JAP2020, Smejkal_PRX2022_2, Mazin_PRX2022}, and in 2022 the name, ``altermagnet'', has been coined by Smejkal {\it et al.} in Refs.~\cite{Smejkal_PRX2022, dwave}.
Importantly, the spin splitting and its associated spin current generation which do not require SOC and the AHE owing to SOC are different in origin and therefore independent to each other (Table I). As we will see below, the conditions for their emergence are not equal; there are cases where the spin-current generation is expected but the AHE is not, and vise versa, cases where the AHE is expected even without the spin splitting. 

Here we should note that, prior to such recent progresses, 
 some of the essential properties of altermagnets have been pointed out previously 
 based on first-principles calculations: spin-split band structure without SOC in collinear antiferromagnets 
by Noda {\it et al.}~\cite{Noda_PCCP2016} and Okugawa {\it et al.}~\cite{Okugawa_JPCM2018}, 
and AHE (at finite frequency) based on collinear antiferromagnetism by Solovyev~\cite{Solovyev_PRB1997}.  
In fact, the rutiles and perovskites were studied as their platforms, 
 in Ref.~\cite{Noda_PCCP2016} and Refs.~\cite{Okugawa_JPCM2018,Solovyev_PRB1997}, respectively. 
The recent novel phenomenon is the prediction of spin-current conductivity~\cite{Ahn_PRB2019,Naka_NatComm2019}
 which can be induced by an electric field, sometimes called electrical spin splitter~\cite{GonzalezHernandez_PRL2021}, 
 or by a thermal gradient~\cite{Naka_NatComm2019}. 
Its schematic illustration in the perovskite structure is shown in Fig.~\ref{fig1}(c).  
\begin{table*}
\centering
\caption{
Microscopic ingredients required for spin splitting, spin current generation and AHE in altermagnets.
The symbol ``$\checkmark$'' (``$-$'') represents ``required'' (``not required'').
}
\vspace{2mm}

\begin{tabular}{cccc}
\hline
& 
\begin{tabular}{c}
AFM $\times$ sublattice \\ (TR symmetry breaking)
\end{tabular}
& 
\begin{tabular}{c}
sublattice-dependent \\ anisotropic electron hopping 
\end{tabular}
&
spin-orbit coupling \\

\hline
\hline

non-relativistic spin splitting & $\checkmark$ & $\checkmark$ & $-$\\
spin current & $\checkmark$ & $\checkmark$ & $-$\\
AHE & $\checkmark$ & $-$ & $\checkmark$ \\

\hline
\end{tabular}
\label{table1}
\end{table*}

In this Review, we shed a light on theoretical progresses on altermagnetic perovskites. 
First, in sec.~\ref{sec:early}, we briefly review the early works mentioned above. 
Next, in sec.~\ref{sec:abo3}, we introduce the crystal structure of perovskites {\it ABX}$_3$ 
 and the canonical microscopic model, i.e., the multiband Hubbard model, 
 to theoretically study the properties of TM-based perovskites 
 with {\it B} sites occupied by $3d$ TM elements. 
In sec.~\ref{sec:ctype}, 
 the AFM ordering that we mainly discuss here as a typical altermagnetic state in perovskites is explained: 
 the {\it C}-type AFM state frequently realized in the $(3d)^2$ case with 2 electrons per TM site. 
In secs.~\ref{sec:splitting} and~\ref{sec:ahe}, 
 we summarize theoretical analyses on the non-relativistic effects and the SOC effects, respectively, 
 based on the model we introduced in sec.~\ref{sec:abo3}. 
We stress here that vast majority of the theoretical studies on altermagnets up to now 
 are based on either first-principles calculations, group theoretical considerations, or simplified model analyses. 
In contrast, an advantage of studying a realistic effective model is that we can simultaneously pin down the microscopic mechanism 
 of each property and directly give feedback to experiments. 
We discuss candidate materials and also recent developments in sec.~\ref{sec:candidate}.
Section~\ref{sec:summary} is devoted to the summary of this review.

\section{Early theoretical work\label{sec:early}}
\begin{figure}[t]
\begin{center}
\includegraphics[width=0.7\columnwidth, clip]{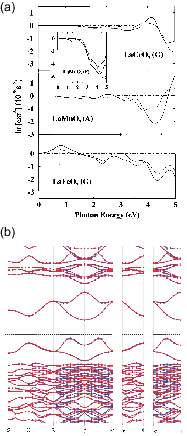}
\end{center}
\caption{
(a) Optical Hall conductivity at finite $\omega$ in various canted AFM states for LaCrO$_3$, LaMnO$_3$, and LaFeO$_3$ 
in Ref.~\cite{Solovyev_PRB1997} 
and (b) spin-split band structure for the $A$-type AFM state in LaMnO$_3$ in Ref.~\cite{Okugawa_JPCM2018}, 
obtained by first-principles calculations.
In (a), $A$, $G$, and $F$-type are conventional notations for the spin ordering patterns on the perovskite structure (see Sec.~\ref{sec:ctype}).
Reprinted (adapted) with permission from Ref.~\cite{Solovyev_PRB1997} (copyright 1997 American Physical Society) and from Ref.~\cite{Okugawa_JPCM2018} (copyright 2018  IOP Publishing).
}
\label{fig2}
\end{figure}

In a pioneering work in Ref.~\cite{Solovyev_PRB1997} more than 25 years ago, 
 the AHE at finite frequency, i.e., the magneto-optical effect, 
 was pointed out to be 
activated in the canted AFM ordered states of the perovskites. 
The optical Hall conductivity $\sigma_{\mu\nu}$ ($\mu \neq \nu$) was calculated 
within first-principles density functional theory  
in the insulating La{\it B}O$_3$ ({\it B} = Cr, Mn, and Fe) with orthorhombic structure, 
i.e., under the GdFeO$_3$-type distortion that we will discuss in detail in the next section.  
Figure~\ref{fig2}(a) shows the antisymmetric components of $\sigma_{\mu\nu}(\omega)$ in the four kinds of spin-ordered patterns that are compatible with the crystal structure. 
Among them, three states accompanied by weak canting FM moments exhibit the AHE, while the remaining one without weak ferromagnetism does not. 
Although  apparently 
the weak ferromagnetism is the origin of the AHE here, 
their magnitudes  
are comparable to that obtained in the FM state shown in the inset, 
and it is concluded that the main AFM components 
are essential. 
We will discuss this point further in sec.~\ref{sec:ahe}. 

More recently, in a series of papers in Refs.~\cite{Noda_PCCP2016,Okugawa_JPCM2018}, 
 the non-relativistic spin-split band structures in MnO$_2$ including its rutile phase and perovskites, respectively, 
 were pointed out within first-principles band calculations, 
 and their symmetry conditions were discussed in detail. 
Ref.~\cite{Okugawa_JPCM2018} 
 in fact investigates the three perovskite compounds same with Ref.~\cite{Solovyev_PRB1997}, 
 in the AFM ordered states; 
 we show in Fig.~\ref{fig2}(b) an example of the spin-split band structure. 
In all of these studies, 
 the importance of the crystal structure underlying the AFM order has been stressed. 
In the next section we will introduce its characteristics and the microscopic model incorporating such ingredients in the perovkite {\it ABX}$_3$ systems. 

\begin{figure}[t]
\begin{center}
\includegraphics[width=1.0\columnwidth, clip]{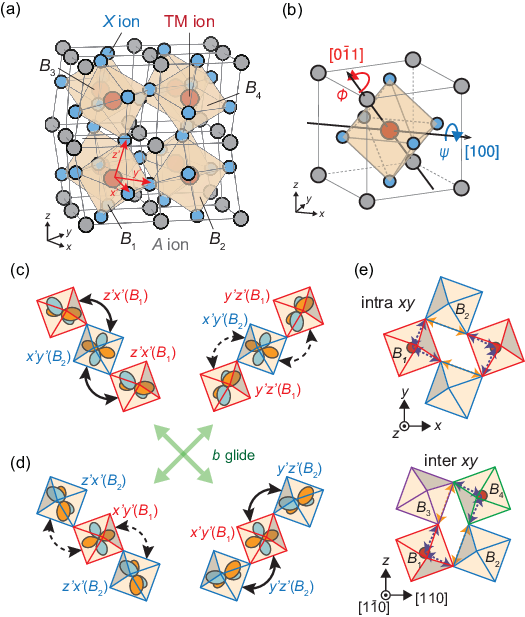}
\end{center}
\caption{
(a) Perovskite structure with the GdFeO$_3$-type distortion. 
$B_1$-$B_4$ denote the {\it BX}$_6$ octahedra contained in the unit cell, connected by symmetry operations thus crystallographically equivalent. 
The $x'y'z'$ axes represent the local coordinate defined for each octahedron.
(b) Two kinds of {\it BX}$_6$ rotation modes of the GdFeO$_3$-type distortion described in the text.  
Schematic illustrations of the anisotropies of the $t_{2g}$-$t_{2g}$ transfer integrals on (c) $B_1$-$B_2$-$B_1$ and (d) $B_2$-$B_1$-$B_2$ bonds along the $[110]$ and $[\bar{1}10]$ directions. 
The magnitude of the transfer integrals denoted by the solid arrows are larger than those by the dashed arrows in the presence of the distortion.
(e) The real-space hopping paths of the intra- and inter-$xy$-plane NNN bonds between the transition metal {\it B} sites through the ligands {\it X}, essential for the AHE.
}
\label{fig3}
\end{figure}

\section{Perovskite structure and model\label{sec:abo3}}
\begin{figure*}[t]
\begin{center}
\includegraphics[width=1.9\columnwidth, clip]{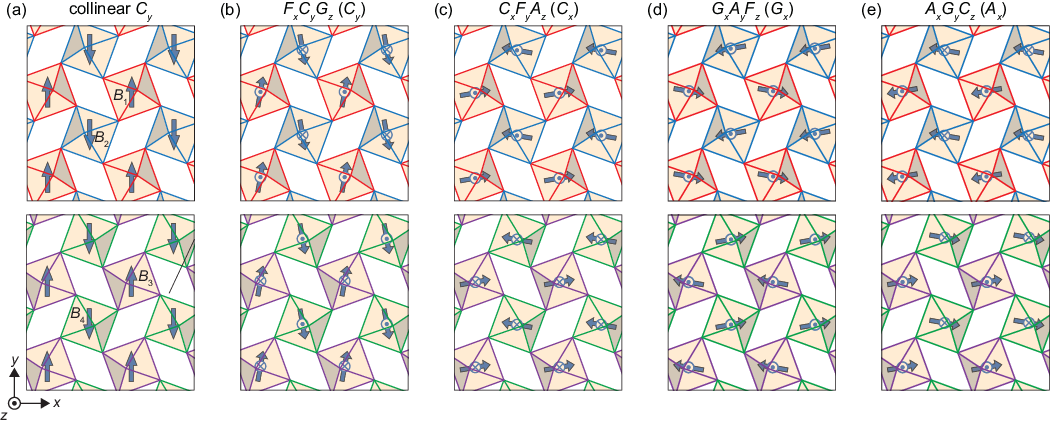}
\end{center}
\caption{
(a) Schematic illustration of the collinear {\it C}-type AFM spin configuration in the two $xy$ planes at $z=\frac{1}{2}$ (upper panel) and $z=0$ (lower panel).
The blue arrows represent up and down spin moments of the $d$ orbitals in the TM ions. 
Here the spins are pointing in the $y$ direction but without SOC their directions can be arbitrarily taken. 
(b) The canted AFM structure with major {\it C}$_y$ component, compatible with the orthorhombic perovkites 
 under the GdFeO$_3$-type distortion, denoted as {\it F}$_x${\it C}$_y${\it G}$_z$ (see text);  
the other canted AFM structures are shown in (c) {\it C}$_x${\it F}$_y${\it A}$_z$, (d) {\it G}$_x${\it A}$_y${\it F}$_z$, and (e) {\it A}$_x${\it G}$_y${\it C}$_z$ with major components along the $x$ axis,
 {\it C}$_x$, {\it G}$_x$, and {\it A}$_x$, respectively.
}
\label{fig4}
\end{figure*}
The perovskite structure consists of {\it BX}$_6$ octahedra sharing corners to form a three-dimensional framework and {\it A} ions occupying the interstitial spaces. 
The important structural feature which makes perovskite system an altermagnet is the rotational distortion of the {\it BX}$_6$ octahedron called GdFeO$_3$-type distortion, commonly seen in many of {\it ABX}$_3$ compounds.
This has conventionally been known and utilized to control the bandwidth via the tolerance factor, a ratio between the ionic radii of three elements.

The distorted structure of {\it ABX}$_3$ is shown in Fig.~\ref{fig3}(a), in which the regularly aligned {\it BX}$_6$ octahedra on the cubic lattice rotate so as to fill the crystal voids around the {\it A} sites, as illustrated in Fig.~\ref{fig3}(b). 
The distortion is parametrized by the major rotation angle $\pm\phi$; the additional tilting $\psi$ is uniquely determined by $\phi$~\cite{OKeeffe_ActaCryst1977}.
Consequently, the crystal becomes orthorhombic and four independent {\it BX}$_6$ octahedra ({\it B}$_1$-{\it B}$_4$) appear within the unit cell.
In Fig.~\ref{fig3}, the global axes, $xyz$, correspond to the crystallographic axes, $abc$, in the space group {\it Pbnm}, or, $cab$, in terms of the equivalent {\it Pnma}. 
The four octahedra are connected by the $b$-glide operation perpendicular to the $x$ axis, the $n$-glide openration perpendicular to the $y$ axis, and the mirror operation perpendicular to the $z$ axis in {\it Pbnm}.

A standard effective model widely used to describe the electronic states of perovskites is the multi-$d$-orbital Hubbard model, 
written as ${\cal H}_{0} + {\cal H}_{\rm int}+ {\cal H}_{\rm SOC}$ including the SOC effect.
The first term describes the electronic hoppings between the $d$ orbitals given by 
\begin{eqnarray} 
{\cal H}_{\rm 0} &=& 
 \sum_{ij \beta \beta' \sigma}^\textrm{NN} [\hat{t}^{dpd}_{ij}(\phi)]_{\beta \beta'} c^{\dagger}_{i \beta \sigma} c_{j \beta' \sigma} \notag \\ 
 &\ & \hspace{3em} + \sum_{ij \beta \beta' \sigma}^\textrm{NNN} [\hat{t}^{dppd}_{ij}(\phi)]_{\beta \beta'} c^{\dagger}_{i \beta \sigma} c_{j \beta' \sigma}, \label{TB}
\end{eqnarray}
where $c_{i \beta \sigma}$ and $n_{i \beta \sigma}(=c^{\dagger}_{i \beta \sigma}c_{i \beta \sigma})$ are the annihilation and the number operators of an electron on TM site $i$ with spin $\sigma$ of the $d$ orbital $\beta$($=x'^2-y'^2, 3z'^2-r^2, x'y', y'z', z'x'$), respectively, represented in the local $x'y'z'$ axes fixed on the $i$th octahedron as shown in Fig.~\ref{fig3}(a).
Here we consider not only the nearest neighbor (NN) but also the next-nearest neighbor (NNN) 
 $d$-$d$ transfer integrals, $\hat{t}^{dpd}_{ij}(\phi)$ and $\hat{t}^{dppd}_{ij}(\phi)$, respectively, 
 evaluated by the hopping processes through the ligand $p$ orbitals~\cite{Naka_PRB2021,Naka_PRB2022}. 
As we will see later, the NNN terms are crucial for the AHE, whose example of hopping paths is shown in Fig.~\ref{fig3}(e). 
Reflecting the {\it BX}$_6$ rotations, the transfer integrals depend on the distortion angle $\phi$. 

An important feature of the NN $d$-$d$ transfer integrals is their spatial anisotropy depending on the bond directions, owing to the hybridization between the different $d$ orbitals induced by the {\it BX}$_6$ rotations. 
As shown in Fig.~\ref{fig3}(c), for example, the inter-orbital transfer integral between $z'x'$ in $B_1$ [$z'x'(B_1)$] and $x'y'$ in $B_2$ [$x'y'(B_2)$] in the $[\bar{1}10]$ direction becomes nonzero, which is zero in the undistorted cubic structure, and is larger than that between $y'z'(B_1)$ and $x'y'(B_2)$ in the $[110]$ direction.
On the other hand, on the $B_2$-$B_1$-$B_2$ bonds, the magnitudes of the electron hoppings in the $[110]$ and $[\bar{1}10]$ directions are switched with each other as shown in Fig.~\ref{fig3}(d), reflecting the $b$-glide symmetry which connects the $B_1$ and $B_2$ sites. 
These inter-orbital hybridizations yield the anisotropic transfer integrals between the same sublattice sites, depending on the bond directions within the $xy$ plane.

The onsite Coulomb interactions on $d$ electrons are introduced in the conventional manner as 
\begin{eqnarray}
{\cal H}_{\rm int} 
&=& U \sum_{i \beta} n_{i \beta \uparrow} n_{i \beta \downarrow} + \frac{U'}{2} \sum_{i \beta \neq \beta'} n_{i \beta} n_{i \beta'} \notag \\
&+& J \sum_{i \beta > \beta' \sigma \sigma'} c^\dagger_{i \beta \sigma} c^\dagger_{i \beta' \sigma'} c_{i \beta \sigma'} c_{i \beta' \sigma} \notag \\
&+& I \sum_{i \beta \neq \beta'} c^\dagger_{i \beta \uparrow} c^\dagger_{i \beta \downarrow} c_{i \beta' \downarrow} c_{i \beta' \uparrow},
\end{eqnarray}
where $U$ and $U'$ represent the intra- and inter-orbital Coulomb interactions, respectively, $J$ is the Hund coupling, and $I$ is the pair hopping interaction. 
The SOC  can be described as 
\begin{eqnarray}
{\cal H}_{\rm SOC} = \zeta \sum_{i} {\bm l}_{\rm loc} \cdot {\bm s}_{\rm loc}, 
\end{eqnarray}
where the $d$ orbital and spin angular momenta are written as 
 ${\bm l}_{\rm loc}$ and  ${\bm s}_{\rm loc}$, respectively, 
 defined in the local $x'y'z'$ axes on the $i$th octahedron.
\begin{figure*}
\begin{center}
\includegraphics[width=1.9\columnwidth, clip]{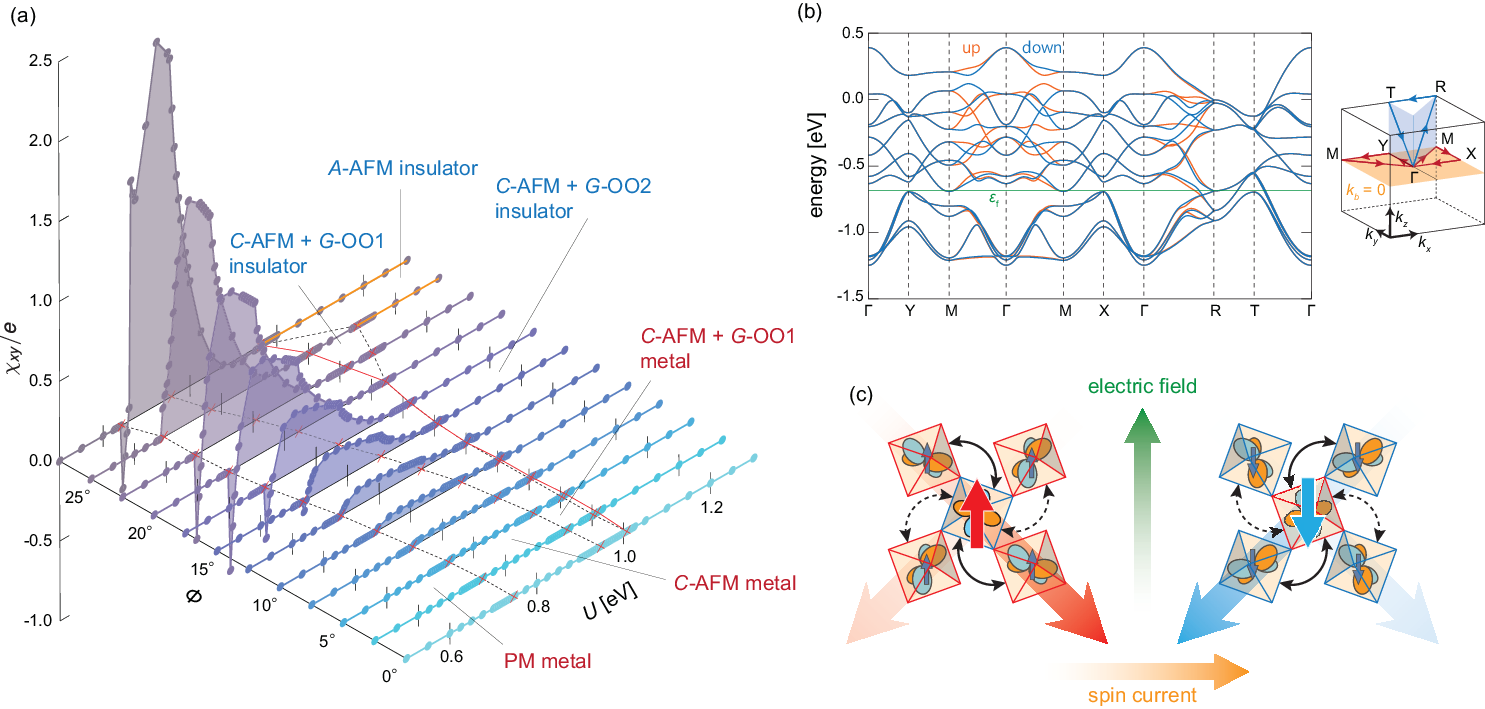}
\end{center}
\caption{
Spin current conductivity $\chi_{xy}$ on the $U$-$\phi$ plane, for the multi-band Hubbard model without SOC. 
The black broken lines on the basal plane indicate the phase boundaries between the different phases. 
The red solid line stands for the MI transition.
(b) Energy band structure of $t_{2g}$ orbitals in the {\it C}-AFM metallic phase at $(U, \phi)=(0.725 \ {\rm eV}, 25^\circ)$.
The Fermi energy is denoted by $\varepsilon_{\rm f}$.
The right panel shows the symmetric lines in the first Brillouin zone.
(c) A schematic view of the spin current generation in altermagnetic perovskites. 
}
\label{fig5}
\end{figure*}
  
\section{C-type antiferromagnetic state\label{sec:ctype}}
To explain the vast variety of electronic properties of perovskites, 
 the multi-$d$-orbital Hubbard model and their variants have been treated extensively~\cite{imada,maekawa}. 
Here we refer to a series of work by Mizokawa and Fujimori~\cite{Mizokawa_PRB1995,Mizokawa_PRB1996,Mizokawa_PRB1999},  
who investigated the spin and orbital ordered phases in $3d$ TM-based perovskite oxides based on the multi-band $d$-$p$ model 
 by varying the electron filling factor, namely, corresponding to different TM elements, 
 providing a systematic view of these materials. 
As for the spin orderings, the basic patterns that they studied are called  
 {\it F}-, {\it A}-, {\it C}-, and {\it G}-types. 
 {\it F} is the ferromagnetic pattern and the other three are AFM states, 
 with {\it A}: in-plane ferromagnetic and inter-plane AFM, {\it C}: in-plane AFM and inter-plane ferromagnetic, 
 and {\it G}: both in-plane and inter-plane AFM patterns. 
In the following two sections, we mainly focus on the {\it C}-type AFM state, 
 typically stabilized in the case where there exist two $d$ electrons per TM element, i.e., the $(3d)^2$ case. 
It shows characteristic altermagnetic properties, 
 and also can be viewed as an analog of the altermagnetism in the $\kappa$-type organic compounds~\cite{Naka_NatComm2019, Naka_PRB2020}. 

Without SOC, the spin direction is not fixed so we can consider the spin-ordered states as the limit from finite SOC to $\zeta=0$. 
In this case 
spin patterns are collinear; a schematic view of the  {\it C}-type AFM state in perovskites are shown in Fig.~\ref{fig4}(a). 
On the other hand, when $\zeta \neq 0$, the spins point to certain directions owing to the magnetic anisotropy and show small canting via the Dzyaloshinsky-Moriya interation. 
Under the space group ({\it Pbnm}\ /\ {\it Pnma}), all the possible AFM spin patterns fall into either of four types~\cite{Treves_PR1962, Bertaut_ActaCryst1968}.
For example, as shown in Fig.~\ref{fig4}(b)
when the $y$-axis spin components align in a {\it C}-type AFM pattern, projections to the other components must show ferromagnetic ({\it F}) moments along the $x$ axis 
 and a {\it G}-type AFM pattern along the $z$ axis; it is written as  {\it F}$_x${\it C}$_y${\it G}$_z$.  
The others are  {\it C}$_x${\it F}$_y${\it A}$_z$, {\it G}$_x${\it A}$_y${\it F}$_z$, and {\it A}$_x${\it G}$_y${\it C}$_z$, as shown in Figs.~\ref{fig4}(c)-\ref{fig4}(e). 
Indeed, these four patterns are the ones investigated by Solovyev~\cite{Solovyev_PRB1997}, 
 who found that the first three are AHE active whereas in the {\it A}$_x${\it G}$_y${\it C}$_z$ state AHE disappears.

\section{Spin current generation and spin-splitting\label{sec:splitting}}

Now we discuss the altermagnetic characters without SOC ($\zeta=0$). 
Figure~\ref{fig5} shows the conductivity of the spin current along the $y$ axis under an electric field along the $x$ axis ($\chi_{xy}$), 
obtained by the Boltzmann transport theory and Hartree-Fock (HF) approximation for ${\cal H}_0 + {\cal H}_{\rm int}$ without the NNN terms, for the $(3d)^2$ case~\cite{Naka_PRB2021}. 
It is plotted as a function of the Coulomb interaction $U$ and the {\it BX}$_6$ rotation angle $\phi$, together with the ground-state phase diagram on the basal plane, 
 choosing model parameters typical of perovskite $3d$-TM oxides. 

In the absence of the GdFeO$_3$-type distortion ($\phi=0$), with increasing $U$, 
 successive phase transitions occur from the paramagnetic (PM) metallic phase, the {\it C}-AFM metallic phase,
and to the {\it C}-AFM insulating phase, accompanied 
by either of the two kinds of {\it G}-type orbital ordering [{\it G}-OO1 and {\it G}-OO2 in Fig.~\ref{fig5}(a)].
Near the transition between the {\it G}-OO1 and {\it G}-OO2 phases, the system undergoes a MI transition. 
These {\it C}-AFM phases are robust against the increase of $\phi$ except for $\phi \gtrsim 25^\circ$ in the large $U$ region, where the {\it A}-type AFM order is stabilized instead.
As shown in Fig.~\ref{fig5}(a), $\chi_{xy}$ is constantly zero at $\phi=0$, while it turns nonzero in the presence of the GdFeO$_3$-type distortion in the {\it C}-AFM metallic phases.

This spin current conductivity is a consequence of the non-relativistic spin splitting in the band structure. 
The energy band structure in the {\it C}-AFM phase is shown in Fig.~\ref{fig5}(b), 
 for $(U, \phi) = (0.775 \ {\rm eV}, \ 25^\circ)$, where the Fermi energy resides in the $t_{2g}$ bands.
One can see the spin splitting in general $\boldsymbol{k}$-points except for the planes
$k_x = 0, \pm 0.5$ and $k_y = 0, \pm 0.5$ in the Brillouin zone. 
This is owing to the TR symmetry breaking by the collinear AFM order and the GdFeO$_3$-type distortion.

In Ref.~\cite{Naka_PRB2021}, by analyzing the group velocities along the spin-split Fermi surfaces for each spin, 
 it is shown that the up-spin and down-spin electrons have different anisotropies in the directions in which they tend to move,  
 i.e., along the $[1\bar{1}0]$ and the $[110]$ directions, respectively.  
As a result of this spin-dependent anisotropy, 
 when an electric field is applied along the $[010]$ direction, 
 the spin current flows along the $[100]$ direction perpendicular to the electric field. 
These spatial anisotropies originate from the GdFeO$_3$-type distortions as mentioned above 
 [see Figs.~\ref{fig3}(c) and \ref{fig3}(d)]. 
When the up-spin and down-spin occupy different sublattices, i.e., $B_1$ and $B_2$, or, $B_3$ and $B4$, 
respectively, owing to the sublattice-dependent anisotropic transfer integrals a finite spin current is induced, 
 whose real-space image is summarized in Fig.~\ref{fig5}(c). 

Let us note the analogy of the spin current generation and the spin splitting here to 
 those first proposed in the organic antiferromagnets $\kappa$-(ET)$_2${\it X}~\cite{Naka_NatComm2019}. 
In $\kappa$-(ET)$_2${\it X}, the anisotropic transfer integrals originate from two kinds of molecules (sublattices) in the unit cell with different orientations, 
 playing the role of the ligands with {\it BX}$_6$ rotations in perovskites. 
Since only one molecular orbital per ET molecule plays a role near the Fermi energy in these quasi-two-dimensional compounds, 
 this anisotropy directly results in the $d$-wave-type spin-splitting in the band structure 
 and explains the mechanism of the spin current generation in a straightforward way. 
In both cases of the perovskites and the organics, 
 spin current conductivity tensor is symmetric within the $xy$-plane,
$\chi_{xy} = \chi_{yx}$, with vanishing diagonal elements, $\chi_{xx} = \chi_{yy} =0$,
 resulting in a peculiar electric-field dependence distinct from the spin Hall effect~\cite{Murakami_Science2003, Sinova_PRL2004}. 
Another difference is that the present spin current generation is a dissipative phenomenon as in the electrical conduction; therefore the temperature variation of the spin current conductivity below the N\'{e}el temperature is expected to be scaled by that of the longitudinal electrical conductivity.

\begin{figure}[t]
\begin{center}
\includegraphics[width=1.0\columnwidth, clip]{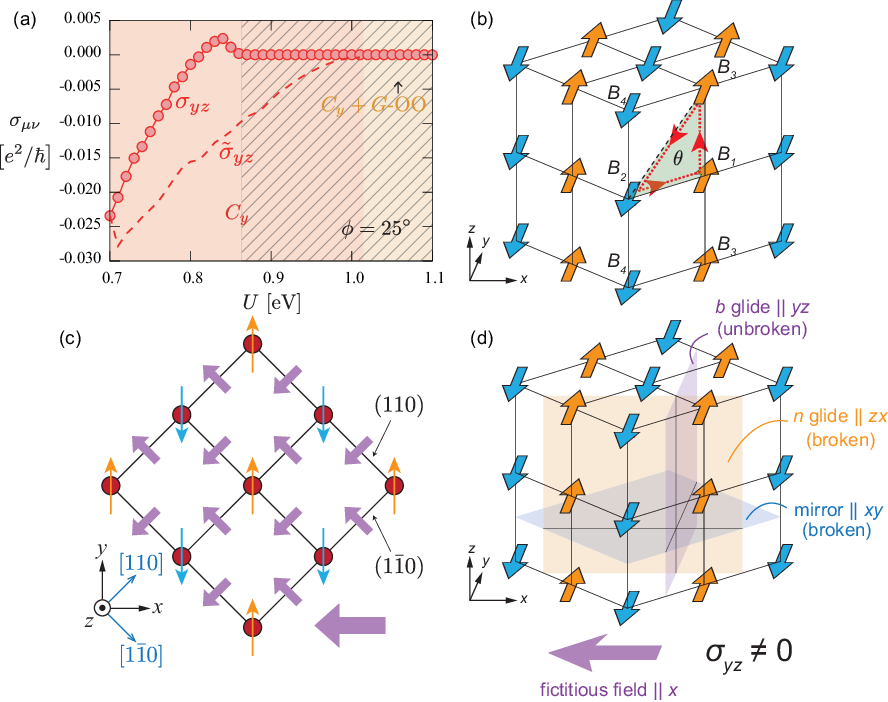}
\end{center}
\caption{
(a) $U$ dependence of the dc Hall conductivity $\sigma_{yz}$ in the $C_y$ phase for $\phi=25^\circ$. 
The hatched area is insulating. $\tilde{\sigma}_{yz}$ are the results obtained by artificially restricting the mean-fields to the major collinear $C_y$ order. 
(b) Schematic view of the  {\it C}$_y$-type collinear AFM state, and an example of the triangular path with fictitious magnetic flux $\theta$. 
(c) The top view of the magnetic flux distribution in the {\it C}$_y$-type AFM state. 
The small purple arrows represent the fictitious magnetic fluxes penetrating the square plaquettes on the $(1\bar{1}0)$ and $(110)$ planes and the large one is the net flux along the $-x$ direction. 
(d) The broken/unbroken symmetries in the  {\it C}$_y$-type AFM state, resultant fictitious field, and active AHE. 
}
\label{fig6}
\end{figure}
\section{Anomalous Hall effect\label{sec:ahe}}

Next, we include the SOC and consider the total Hamiltonian, ${\cal H}_{\rm 0} + {\cal H}_{\rm int} + {\cal H}_{\rm SOC}$, 
 to discuss the AHE, the condition for its appearance, and its microscopic mechanism~\cite{Naka_PRB2022}. 
Within HF approximation to a three-band model for the $t_{2g}$ electrons, 
 similarly to the case without the SOC [Fig.~\ref{fig5}(a)], the {\it C}-type AFM states are most stable over a wide parameter range in the $(3d)^2$ case but with the spin directions fixed and not purely collinear 
  as introduced in sec.~\ref{sec:ctype}. 
The spin pattern is predominantly  {\it C}$_x$-type AFM ($C_x F_y A_z$) [Fig.~\ref{fig4}(c)], with moments mostly oriented along the $x$-axis when $\phi$ is small. 
On the other hand, in the large $\phi$ region, the $C_y$-type AFM state ({\it F}$_x${\it C}$_y${\it G}$_z$) with the moments mostly along the $y$-axis becomes stable [Fig.~\ref{fig4}(b)].
For simplification, we will represent each pattern by the major component with the largest projected spin moments, e.g., 
 as {\it C}$_y$ pattern. 
In the following, we focus on the AHE in the {\it C}$_y$ phase stabilized in the large $\phi$ region.

Figure~\ref{fig6}(a) shows the $U$ dependence of the dc Hall conductivity $\sigma_{yz}$ in the $C_y$ phase for $\phi = 25^\circ$.
Since there is the ferromagnetic component $F_x$ and the AHE appears in the plane perpendicular to its direction, one might recall the case for conventional ferromagnets. 
However, their magnitude compared to the net moment is exceptionally large.
In fact, when we retain only the major component $C_y$ by artificially restricting the mean-field solutions to the collinear $C_y$ order as illustrated in Fig.~\ref{fig4}(a), the calculated Hall conductivity $\tilde{\sigma}_{yz}$, also plotted in Fig.~\ref{fig6}(a), roughly recovers the original $U$ dependence (it is actually larger than $\sigma_{yz}$).
This trend is also seen in $\sigma_{zx}$ and $\tilde{\sigma}_{zx}$ in the $C_x$ phase, 
 where the ferromagnetic moment is parallel to the $y$ axis.
These indicate that the collinear AFM order is essential for the AHE. 

Similarly to the case of spin splitting discussed in the previous section, 
the AHE vanishes at $\phi=0$, 
 laying out the necessity of the GdFeO$_3$-type distortion. 
 On the other hand, the $\phi$ dependences of $\chi_{xy}$ in Fig.~\ref{fig5}(a) and that of $\sigma_{yz}$ in Ref.~\cite{Naka_PRB2022} are qualitatively distinct; $\chi_{xy}$ increases while $\sigma_{yz}$ decreases with increasing $\phi$.
Furthermore, we should note that the AHE always disappears when we set the NNN terms to zero, even for finite $\phi$. 
These demonstrate the crucial difference
 between the AHE and the nonrelativistic spin splitting with its resultant spin current generation, 
 since the latter do not require SOC nor the NNN terms.  
We will discuss their roles from a microscopic viewpoint in the following. 

A microscopic interpretation of the AHE by the real space distribution of the fictitious magnetic field acting on conduction electrons 
 provides an intuitive understanding, as discussed in non-collinear magnets such as in the kagome lattice~\cite{Ohgushi_PRB2000,Tomizawa_PRB2009}.
It is useful in the collinear (or canted) AFM cases as well, as demonstrated first in the case of $\kappa$-(ET)$_2${\it X}~\cite{Naka_PRB2020}, 
and also in the perovskites~\cite{Naka_PRB2022}.  
As shown in Fig.~\ref{fig6}(b), we take the {\it C}$_y$ pattern as an example; now knowing that NNN electron hopping is essential for the AHE,
 let us consider the smallest triangular paths consists of two NN bonds and one NNN bond (an example is shown in the figure).  
The magnetic flux $\theta$ passing through this triangle $ijk$ is written as $\theta = \arg(t_{ij}t_{jk}t_{ki})$, 
 where $t_{\mu\nu}$ ($\mu, \nu = i, j, k$) are the transfer integrals between sites $\mu$ and $\nu$, 
 which generally become complex numbers due to the SOC and contribute to $\theta$. 
In each plaquette, there are four possible triangular paths, 
 which cancel out in the paramagnetic state, but do not when the AFM order takes place and provide a finite fictitious magnetic field; 
 the non-equivalent paths with excess up or down spins are responsible for this unbalance. 
By counting them for all the plaquettes, one obtains a distribution as shown in Fig.~\ref{fig6}(c), 
 which results, by summing them up, in a net fictitious field along the $x$-direction and the $y$ component cancels out. 
Therefore, the electrons moving under an electric field in the $yz$ plane drift perpendicular to both the electric field and the fictitious field, resulting in the AHE. 

\begin{table}
\centering
\caption{
Relation between the AHE and the symmetries of the AFM patterns on {\it Pbnm} structure. 
Y (N) represents the presence (absence) of the symmetry.}
\vspace{2mm}
\begin{tabular}{ccccc}
\hline
AFM & mirror $\perp$ $z$ & $b$ glide $\perp$ $x$ & $n$ glide $\perp$ $y$ & AHE \\
\hline
\hline
$C_xF_yA_z$ & N  & N & Y & $\sigma_{xz}$ \\
$F_xC_yG_z$ & N  & Y & N & $\sigma_{yz}$ \\
$G_xA_yF_z$ & Y  & N & N & $\sigma_{xy}$ \\
$A_xG_yC_z$ & Y  & Y & Y & N/A \\
\hline
\end{tabular}
\label{table2}
\end{table}
Finally, we discuss the relationship between the symmetry of the four possible AFM patterns discussed in Sec.~\ref{sec:ctype} and the AHE. 
As mentioned above, 
 there are three independent symmetry operations: a mirror operation perpendicular to the $z$-axis, a glide operation in the $y$-direction concerning a plane perpendicular to the $x$ axis ($b$ glide), and a glide operation in the $x+z$ direction concerning a plane perpendicular to the $y$-axis ($n$ glide).
Table~\ref{table2} summarizes the relationship between the broken symmetries 
and the resultant AHE. 
When two of the three symmetries are broken by AFM ordering, a fictitious magnetic field arises in the direction of the intersection line of these two mirror/glide planes, and an AHE occurs in the plane perpendicular to this direction; 
 in the $A_xG_yC_z$ phase, where all the three symmetries are preserved, the AHE vanishes. 
This is consistent with the early {\it ab initio} study shown in Fig.~\ref{fig2}(a)~\cite{Solovyev_PRB1997}. 
Additionally, this rule is also applicable to the organic altermagnet $\kappa$-(ET)$_2X$ belonging to the same space group $Pbnm$, where the AHE discussed in Ref.~\cite{Naka_PRB2020} corresponds to $G_xA_yF_z$ in this table. 
Note that the $s^x$, $s^y$, and $s^z$ components that constitute a given AFM pattern shown in Table~\ref{table2} all have the same symmetry as each other. 
Taking the  {\it F}$_x${\it C}$_y${\it G}$_z$ as an example, each component  {\it F}$_x$, {\it C}$_y$, and {\it G}$_z$  
 breaks the mirror and $n$ glide symmetries even when one of them exists independently, 
 for example shown as in Fig.~\ref{fig6}(d) for the collinear {\it C}$_y$ pattern discussed above. 
Thus, also from the viewpoint of symmetry, the AHE can occur in the presence of a single collinear AFM component, 
 even in the absence of the ferromagnetic net moment. 
\begin{table*}
\centering
\caption{
A list of candidates of altermagnetic perovskites, their AFM order, and expected cross-correlation phenomena with the $d$ electron configurations, N\'{e}el temperatures, and the $B$-$X$-$B$ bond angles, whose deviation from 180$^\circ$ indicates the degree of the GdFeO$_3$-type distortion. 
The bold symbols represent the major compoments of the AFM order. 
In particular, since there are many compounds in the series of {\it A}CrO$_3$ and {\it A}FeO$_3$, only the representatives are listed here; for the other similar candidates, a review in Ref.~[\onlinecite{Bousquet_JPhys2016}] will be helpful.
}
\begin{tabular}{ccccccc}
\hline
$d^n$ & compound & AFM & $T_{\rm N}$ & $B$-$X$-$B$ angle & reference & cross correlation \\
\hline
\hline
$d^1$ & LaTiO$_3$ & ${\bm G_x}A_yF_z$ & $146$ K & $157^\circ$ & [\onlinecite{MacLean_JSSC1971, Cwik_PRB2003}] & AHE ($\sigma_{xy}$) \\
$d^2$ & CaCrO$_3$ & $F_x{\bm C_y}G_z$ & $90$ K & $160^\circ$ & [\onlinecite{Komarek_PRL2008}] & spin current ($\chi_{xy}$), AHE ($\sigma_{yz}$) \\
$d^2$ & LaVO$_3$ & $F_x{\bm C_y}G_z$ & $143$ K & $158^\circ$ & [\onlinecite{Bordet_JSSC1993, Miyasaka_PRB2003}]& spin current ($\chi_{xy}$), AHE ($\sigma_{yz}$) \\
$d^3$ & LaCrO$_3$ & ${\bm G_x}A_yF_z$ or $F_xC_y{\bm G_z}$ & $298$ K & $160^\circ$ & [\onlinecite{Zhou_PRL2011}] & AHE ($\sigma_{xy}$ or $\sigma_{yz}$) \\
$d^3$ & YCrO$_3$ & ${\bm G_x}A_yF_z$ & $142$ K & $148^\circ$ & [\onlinecite{Judin_SSC1966, Tiwari_JPCM2013}] & AHE ($\sigma_{xy}$) \\
$d^3$ & Ca$_{1-x}$La$_x$MnO$_3$ & ${\bm G_x}A_yF_z$ & $110$-$125$ K & $158^\circ$ ($x=0.1$) & [\onlinecite{Wollan_PR1955, Ling_PRB2003, Wang_JPCC2009}] & AHE ($\sigma_{xy}$) \\
$d^3$ & Ca$_{1-x}$Sr$_{x}$MnO$_3$ & ${\bm G_x}A_yF_z$ & $125$-$175$ K & $159$-$167^\circ$ ($x=0$-$0.5$) & [\onlinecite{Kafalas_JAP1971, Chmaissem_PRB2001}] & AHE ($\sigma_{xy}$) \\
$d^3$ & Ca$_{1-x}$Ce$_x$MnO$_3$ & $F_xC_y{\bm G_z}$ & $\sim 115$ K & $158^\circ$ ($x=0.025$-$0.075$) & [\onlinecite{Caspi_PRB2004}] & AHE ($\sigma_{yz}$) \\
$d^4$ & LaMnO$_3$ & ${\bm A_y}F_z$ & $140$ K & $168^\circ$ & [\onlinecite{Skumryev_EPJB1999, Bousquet_JPhys2016}] & AHE ($\sigma_{xy}$) \\
$d^5$ & LaFeO$_3$ & ${\bm G_x}A_yF_z$ & $738$ K & $156^\circ$ & [\onlinecite{White_JAP1969, Bousquet_JPhys2016, Dixon_JSSC2015}] & AHE ($\sigma_{xy}$) \\
$d^5$ & YFeO$_3$ & ${\bm G_x}A_yF_z$ & $644$ K & $151^\circ$ & [\onlinecite{White_JAP1969, Bousquet_JPhys2016, Bharadwaj_CI2021}] & AHE ($\sigma_{xy}$) \\
$d^5$ & NaMnF$_3$ & ${\bm G_x}A_yF_z$ & $66$ K & $141^\circ$ & [\onlinecite{Shane_JAP1967}] & AHE ($\sigma_{xy}$) \\
$d^5$ & KMnF$_3$ & ${\bm G_x}A_yF_z$ & $88$ K & $168^\circ$ & [\onlinecite{Beckman_PR1961, Heeger_PR1961, Knight_JAC2020}] & AHE ($\sigma_{xy}$) \\
\hline
\end{tabular}
\label{table3}
\end{table*}

\section{Candidate materials\label{sec:candidate}}
Now, we identify potential altermagnetic perovskites where the observations of spin current generation and the AHE are expected, which are seen in a wide range of electron configurations ($d^n$) as summerized in Table~\ref{table3}. 
As for the spin current generation, CaCrO$_3$, which contains Cr$^{4+}$ ions with $(3d)^2$ configuration, is one of the plausible candidates.
This compound exhibits {\it C}$_y$-type AFM order below $90$ K and shows metallic conduction even below the N\'{e}el temperature~\cite{Komarek_PRL2008}. 
The {\it C}-type AFM order is also observed in vanadium oxides {\it A}VO$_3$ ({\it A} = La-Y) with $(3d)^2$~\cite{Bordet_JSSC1993, Miyasaka_PRB2003}; 
although this phase is generally insulating accompanied by the {\it G}-type OO, this can be suppressed and metallic under carrier doping by substitution of the {\it A}-site cation~\cite{Miyasaka_PRL2000}. 
Furthermore, the spin current generation mechanism discussed here is not limited to $d^2$ systems but can also be applied to systems exhibiting the {\it C}-type AFM order with different numbers of $d$ electrons. 
For example, in $d^4$ systems, similar spin splitting is realized~\cite{Naka_PRB2021, Okugawa_JPCM2018} owing to $e_g$-$e_g$ and $e_g$-$t_{2g}$  electron hoppings. 
This suggests that manganese oxides containing $(3d)^4$ ions and exhibiting the {\it C}-type AFM, such as {\it R}$_x${\it A}$_{1-x}$MnO$_3$, could also be potential candidates~\cite{Solovyev_PRB1997, Kajimoto_PRB2002}.
Here we focus only on the {\it C}-type AFM order in which metallic states have been experimentally observed. 
However, since the spin splitting occurs also in {\it A}- and {\it G}-type AFM ordered phases~\cite{Okugawa_JPCM2018, Naka_PRB2022}, they could be candidates if they are metallized by carrier doping or proximity effect.

CaCrO$_3$ is also a candidate for the dc AHE, where $\sigma_{yz}$ is expected in the {\it C}$_y$-type AFM metallic phase below $90$ K. 
In fact, the AHE in CaCrO$_3$ has recently been investigated by first-principles calculation by Nguyen {\it et al.}~\cite{Nguyen_PRB2023} and indeed 
 shows the dc Hall conductivity in the $C_x$ and $C_y$ phases 
 which are qualitatively consistent with the model calculation~\cite{Naka_PRB2022}. 
If we consider the ac AHE, the series of {\it A}TiO$_3$ with $(3d)^1$ and {\it A}VO$_3$ with $(3d)^3$ are preferred as well.
For example, LaTiO$_3$ exhibits the $G_x$ phase accompanied by a complex OO state with distorted TiO$_6$ octahedra~\cite{MacLean_JSSC1971, Cwik_PRB2003} and LaVO$_3$ shows the {\it C}$_y$ + {\it G}-OO phase shown in Fig.~\ref{fig6}(a) below $140$ K, where $\sigma_{yz}(\omega)$ at finite $\omega$ is expected~\cite{Naka_PRB2022}.
Other typical examples of $d^3$ system are LaCrO$_3$ and YCrO$_3$ containing Cr$^{3+}$.
Either {\it F}$_x${\it C}$_y${\it G}$_z$ or {\it G}$_x${\it F}$_y${\it A}$_z$ is indicated in experiments~\cite{Zhou_PRL2011} in LaCrO$_3$ and 
{\it G}$_x${\it F}$_y${\it A}$_z$ is confirmed in YCrO$_3$~\cite{Judin_SSC1966, Tiwari_JPCM2013}.
Besides, perovskites containing $A^{2+}$ and Mn$^{4+}$, such as CaMnO$_3$, and its $A$-site substitutions, are also possible candidates.
For example, {\it G}$_x$-type AFM state is realized in Ca$_{1-x}$La$_x$MnO$_3$ with $0 \leq x < 0.07$~\cite{Wollan_PR1955, Ling_PRB2003, Wang_JPCC2009} and Ca$_{1-x}$Sr$_{x}$MnO$_3$ with $0 \leq x < 0.5$~\cite{Kafalas_JAP1971, Chmaissem_PRB2001}, and {\it G}$_z$-type AFM is realized in Ca$_{1-x}$Ce$_x$MnO$_3$ with $0.025 \leq x < 0.075$~\cite{Caspi_PRB2004}; 
their mother compound LaMnO$_3$ with $d^4$ configuration showing the {\it A}$_y$-type AFM is also a canditate~\cite{Skumryev_EPJB1999, Bousquet_JPhys2016}.
Furthermore, LaFeO$_3$ and YFeO$_3$, along with other rare-earth based compounds, which are $(3d)^5$ systems containing Fe$^{3+}$, also exhibit {\it G}$_x$-type AFM state~\cite{White_JAP1969, Solovyev_PRB1997, Bousquet_JPhys2016, Dixon_JSSC2015, Bharadwaj_CI2021}.
Not only oxides but also fluorides, such as NaMnF$_3$~\cite{Shane_JAP1967} and KMnF$_3$~\cite{Beckman_PR1961, Heeger_PR1961, Knight_JAC2020}, are promising candidates.

\section{Summary and perspectives\label{sec:summary}}

In summary, we have reviewed theoretical studies on altermagnetic properties in perovskites.
The spin splitting in the antiferromagnetic ordered phase pointed out in the early first-principles calculations is reproduced by the multi-$d$-orbtial Hubbard model analysis.
The microscopic origin is a cooperative effect of the antiferromagnetic ordering and the spatially anisotropic inter-orbital electron hoppings that depend on the sublattices caused by the GdFeO$_3$-type lattice distortion.
This spin splitting gives rise to the spin current generation characterized by the symmetric conductivity tensor under an external electic field in the antiferromagnetic metallic phase.
Besides, when the relativistic spin-orbit coupling is introduced, the anomalous Hall effect emerges in the canted antiferromagnetic phase, where the next-nearest-neighbor transfer integrals as well as the nearest-neighbor ones play an important role for generating the net fictitious magnetic field.
Although the anomalous Hall effect appears in the plane perpendicular to the weak ferromagnetic moment, the essence is the collinear antiferromagnetic component on the GdFeO$_3$ structure that has the same symmetry with the ferromagnetism.

Researches on altermagnetism is now explosively expanding, represented by the direct observations of the spin-split bands~\cite{Krempasky_Nat2024, Osumi_PRB2024}. 
Other cross-correlation phenomena are discussed, such as piezomagnetic effect~\cite{Aoyama_PRM2024}, cooperative effects with superconductivities~\cite{Sumita_PRR2023}, and topological insulators~\cite{Fernandes_PRB2024}, which lead us to expect further discovery of novel phenomena in the future.
For example, recently an AHE originating from the proximity effect at the interface between two kinds of distorted perovskites, showing a paramagnetic metal and an AFM insulating phases, has been experimentally observed~\cite{Fujita_arXiv2024}. 
This might be interpreted as a cooperative effect of the present altermagnetic AHE and the so-called topological Hall effect owing to the scaler spin chirality~\cite{Ohgushi_PRB2000}, which is an interesting issue awaiting further theoretical analysis.

Perovskites have been synthesized in various forms and numerous antiferromagnetic ordered phases have been studied. 
In contrast, Table~\ref{table3} gives only a partial and non-exhaustive list of representative {\it ABX}$_3$ systems with $B$ taking $3d$-TM elements, and therefore many other compounds not included here, such as Ruddlesden-Popper-type {\it A}$_{n+1}${\it B}$_n${\it X}$_{3n+1}$, and $4d$ and $5d$ perovskites, can also be promising candidates for altermagnetic perovskites. 
Reexamination and reconsideration of their electronic states are crucial for exploring highly functional altermagnets, and further experimental studies in this field are greatly anticipated.
Let us conclude this article hoping that the series of research will pave the way to the realization of ``interesting and applicable antiferromagnets" beyond N\'{e}el's prediction.

\begin{acknowledgments}
The authors would like to thank T. Aoyama, T. C. Fujita, Y. Fuseya, T. Katsufuji, M. Kawasaki, H. Kishida, J. Matsuno, T. Mizokawa, M. Mizumaki, M. Mochizuki, T. P. T. Nguyen, K. Omura, I. V. Solovyev, and K. Yamauchi for valuable comments and discussions. 
This work was supported by Grant-in-Aid for Scientific Research, No. JP19K03723, JP19K21860, JP19H05825, JP20H04463, JP23H01129, JP23K25826, JP23K03333, JST-CREST (No. JPMJCR18T2), the GIMRT Program of the Institute for Materials Research, Tohoku University, No. 202112-RDKGE-0019.
\end{acknowledgments}



\begin{thebibliography}{99}
\bibitem{net_magnetization}
In the presence of the spin-orbit coupling, there are cases where small spin canting appears and net magnetization becomes nonzero.
\bibitem{PT_symmetry}
We note that if the $PT$ symmetry, which is combination of the macroscopic TR symmetry $T$ and space-inversion symmetry $P$, is preserved, the spin splitting is prohibited even when $T$ is broken, and such a system is not classified as an altermagnet~\cite{Smejkal_PRX2022}.
\end{thebibliography}
\end{document}